\documentclass[12pt]{article}

\usepackage{amsfonts}
\usepackage[centertags]{amsmath}
\usepackage{amssymb}
\usepackage{cite}
\usepackage{psfrag}
\usepackage{graphicx}
\usepackage{bbm,bm}
\usepackage{epsfig, multicol}

\allowdisplaybreaks
\begin{document}

\topmargin 0pt
\oddsidemargin 0mm
\def\be{\begin{equation}}
\def\ee{\end{equation}}
\def\bea{\begin{eqnarray}}
\def\eea{\end{eqnarray}}
\def\ba{\begin{array}}
\def\ea{\end{array}}
\def\ben{\begin{enumerate}}
\def\een{\end{enumerate}}
\def\nab{\bigtriangledown}
\def\tpi{\tilde\Phi}
\def\nnu{\nonumber}
\newcommand{\eqn}[1]{(\ref{#1})}

\newcommand{\half}{{\frac{1}{2}}}
\newcommand{\vs}[1]{\vspace{#1 mm}}
\newcommand{\dsl}{\pa \kern-0.5em /} 
\def\a{\alpha}
\def\b{\beta}
\def\g{\gamma}\def\G{\Gamma}
\def\d{\delta}\def\D{\Delta}
\def\ep{\epsilon}
\def\et{\eta}
\def\z{\zeta}
\def\t{\theta}\def\T{\Theta}
\def\l{\lambda}\def\L{\Lambda}
\def\m{\mu}
\def\f{\phi}\def\F{\Phi}
\def\n{\nu}
\def\p{\psi}\def\P{\Psi}
\def\r{\rho}
\def\s{\sigma}\def\S{\Sigma}
\def\ta{\tau}
\def\x{\chi}
\def\o{\omega}\def\O{\Omega}
\def\k{\kappa}
\def\pa {\partial}
\def\ov{\over}
\def\nn{\nonumber\\}
\def\ud{\underline}
%
\begin{center}
{\large{\bf Evidence for non-supersymmetric AdS/CFT}}

\vs{5}

{Shibaji Roy\footnote{E-mail: shibaji.roy@saha.ac.in}}

{\it Saha Institute of Nuclear Physics\\
1/AF Bidhannagar, Kolkata 700064, India\\}

\end{center}

\vs{5} 

\begin{abstract}
AdS/CFT is a conjectured equivalence between a field theory without gravity (conformal field theory) and a string theory
in a special curved background (anti de-Sitter space), where theories on both sides of the equivalence is supersymmetric.
Since nature as we observe at low energy is non-supersymmetric, it will be very useful if we can have this conjecture
in the non-supersymmetric case. Here we will give some evidence that AdS/CFT type duality must work for non-supersymmetric
system. We will indicate some calculation which leads us to such conclusion.  

\end{abstract}

\vs{5}
 
AdS/CFT correspondence, in its original version \cite{Maldacena:1997re} (see also \cite{Witten:1998qj,Gubser:1998bc}), is 
a conjectured equivalence or a duality between 
a non-gravitational, conformally invariant, supersymmetric field theory in four dimensions
(more precisely, $D=4$, ${\cal N}=4$ super Yang-Mills theory) and a string theory (type IIB)
or a gravitational theory in AdS space in five dimensions (times a five-dimensional sphere). It is 
holographic and is a strong-weak duality symmetry, in the sense, that when the field theory is strongly
coupled the string theory is weakly coupled (given by supergravity) and vice-versa. This duality, is
therefore, very useful to understand the strong coupling behavior of field theory (like, for example, QCD
at low energy) by studying the weakly
coupled string theory or supergravity. However, the theories on both sides of this duality are supersymmetric
and therefore not very realistic. For example, on the gravity side, the AdS geometry which is maximally
supersymmetric arise from the near horizon limit of BPS D3 brane solution of type IIB string theory and
is dual to a field theory which has superconformal symmetry \cite{Aharony:1999ti}, unlike the realistic QCD theory.

So, although AdS/CFT correspondence is very successful in the supersymmetric case, there is no doubt that 
it will be more useful if it can be understood for the non-supersymmetric case for the reason we mentioned earlier
that nature in low energy (at least the energy at which today's accelerators operate) is not supersymmetric.
So, in order to understand the strongly coupled field theory like low energy QCD
(which is neither supersymmetric nor conformal),
we must understand AdS/CFT in the non-supersymmetric case. In all the previous works, where strong coupling
behavior of various non-supersymmetric field theoretic systems have been studied using AdS/CFT type correspondence 
(or more generally gauge/gravity correspondence), it has been assumed that AdS/CFT
correspondence must hold good for the non-supersymmetric case. But there was no evidence in favor of this assumption.
Here we will provide some evidence by studying the scattering of a minimally coupled scalar in the background
of non-supersymmetric D3 branes. We will obtain the scattering equation of motion of this scalar (which turns out
to be the same even for the graviton) and obtain the form of the potential the scalar experiences while moving
in the background. Since the scattering equation is quite complicated it is not easy to solve it (with all its
generality) analytically. So, we study the scattering potential closely and plot it with the
radial distance from the brane. We find that in the low energy limit (also called the decoupling limit), as
the scalar (or the graviton) approaches the branes, the potential becomes infinite and therefore, the scalar (or the
graviton) is never able to reach the brane. This clearly indicates that the bulk gravity indeed gets decoupled from
the non-supersymmetric brane and provides a strong evidence that it may be possible to extend AdS/CFT correspondence
for the non-supersymmetric cases. We have even worked out the decoupling limit of non-supersymmetric D3 brane
in certain situation and obtained the throat geometry which serves as the gravity dual of some QCD-like theory.
This geometry has been studied before as dual to some QCD-like theory assuming that AdS/CFT works for the 
non-supersymmetric case and we have provided some basis for their assumptions.         

The non-supersymmetric D3 brane solution of type IIB string theory has the following form \cite{Zhou:1999nm,Brax:2000cf,Lu:2004ms,Nayek:2015tta},
\bea\label{nonsusyd3n}
ds^2 & = & F(\rho)^{-{\frac{1}{2}}}G(\rho)^{\frac{\delta}{4}}\left(-dt^2+\sum_{i=1}^3 (dx^i)^2\right)+F(\rho)^{\frac{1}{2}}
G(\rho)^{\frac{1+\delta}{4}}\left(\frac{d\rho^2}{G(\rho)}+\rho^2 d\Omega^2_5\right)\nn
e^{2\phi} & = & G(\rho)^\delta\nn
F_{[5]} & = & \frac{1}{\sqrt{2}}(1 + \ast) Q {\rm Vol}(\Omega_5)
\eea
where $G(\rho) = 1 + \rho_0^4/\rho^4$ and $F(\rho) =  G(\rho)^{\frac{\alpha}{2}} \cosh^2\theta - G(\rho)^{-\frac{\beta}{2}} \sinh^2\theta$.
In the above the metric is given in the string frame and $\phi$ is the dilaton. Note that we have suppressed the string coupling
constant $g_s$ which is assumed to be small. $F_{[5]}$ is the self-dual Ramond-Ramond 5-form and $Q$ is the charge of the 
non-supersymmetric D3 brane. The solution has a naked singularity at $\rho=0$ and regular everywhere else. Note that the
solution is characterized by six parameters $\a$, $\b$, $\d$, $\theta$, $Q$ and $\rho_0$, of which $\rho_0$ has the dimension
of length, $Q$ has the dimension of four-volume and the rest are dimensionless. However, not all the parameters are
independent as they satisfy certain constraints originating from the consistency of the equation of motion. The constraints 
are \cite{Nayek:2015tta},
\bea\label{constraints}
& & \a = \b\nn
& & \a^2 + \d^2 = \frac{5}{2} \quad \Rightarrow \quad |\d| \leq \sqrt{\frac{5}{2}}\nn
& & Q = 2 \a \rho_0^4 \sinh2\theta
\eea  
Using \eqref{constraints}, we find that the solution is characterized by three independent parameters $\theta$, $\delta$ and $\rho_0$.
The BPS D3 brane solution can be recovered from this non-supersymmetric solution \eqref{nonsusyd3n} if we take a double scaling limit
$\rho_0 \to 0$, $\theta \to \infty$, such that, $(\a/2) \rho_0^4 (\cosh^2\theta + \sinh^2\theta) \to R^4$ = fixed.
Now with this we can easily check that $G(\rho) \to 1$ and $F(\rho) \to {\bar H}(\rho) = 1 + R^4/\rho^4$ and $Q \to 4 R^4$. 
Substituting these in \eqref{nonsusyd3n} the non-supersymmetric D3 brane solution reduces precisely to a BPS D3 brane solution.

Next we couple a scalar $\varphi$ minimally to the background given in \eqref{nonsusyd3n} and the corresponding action will be
\be\label{action}
S = \frac{1}{4\pi G_{10}}\int d^{10}x\sqrt{-g}e^{-2\phi}\partial_\mu\varphi\partial^\mu\varphi
\ee
where $g$ is the determinant of the background metric \eqref{nonsusyd3n}, $\phi$ is the dilaton and $G_{10}$ is the ten dimensional Newton's 
constant. The equation of motion of the scalar following from \eqref{action} is,
\be\label{eom}
D_\mu\partial^\mu \varphi - 2 \partial_\mu\phi\partial^\mu\varphi = 0
\ee
We will assume that $\varphi$ has spherical symmetry in the transverse space (we are considering s-wave) and is independent of
the spatial coordinates of the brane (assumed for simplicity). So, $\varphi$ has the form, $\varphi = \Phi(\rho)e^{i\omega t}$. Substituting
this in \eqref{eom} we get the following second order differential equation,
\be\label{diffeqn}
\partial_\rho^2\Phi(\rho) + \left[\frac{5}{\rho}+\frac{\partial_\rho G(\rho)}{G(\rho)}\right]\partial_\rho \Phi(\rho) + \omega^2 F(\rho)
G(\rho)^{-\frac{3}{4}}\Phi(\rho) = 0
\ee
By redefining the radial coordinate as, $u = \omega\rho$ and introducing a function $g(u)$ by $\Phi(u) = k(u) g(u)$, with $k(u) = 1/\sqrt{u^5 G(u)}$, 
we can rewrite \eqref{diffeqn} in a Scr\" odinger-like form in terms of the new function $g(u)$ as 
follows,
\be\label{schform}
\left(\partial_u^2 - V(u)\right) g(u) = 0
\ee
where
\be\label{potential}
V(u) = \frac{15}{4u^2} - \frac{1}{4} \left(\frac{\partial_u G(u)}{G(u)}\right)^2 - F(u) G(u)^{-\frac{3}{4}}
\ee
and $G(u) = 1 + (\omega \rho_0)^4/u^4$, $F(u) = G(u)^{\a/2} \cosh^2\theta - G(u)^{-\a/2}\sinh^2\theta$. Once we have the form of $g(u)$ by solving
\eqref{schform}, we can obtain the form of $\Phi$ as 
\be\label{Phi}
\Phi(t,\rho) = \frac{1}{(\omega\rho)^{\frac{5}{2}}}\frac{g(\omega\rho)}{\sqrt{G(\omega\rho)}} e^{i\omega t}
\ee 
Eq.\eqref{potential} represents the scattering potential the scalar will see while moving in the background of non-supersymmetric D3 brane.
In fact one can show that graviton also experiences the same potential \cite{Nayek:2015tta} and we will show that either the scalar or 
the graviton does not reach the brane in a low energy limit showing that gravity decouples from the brane. 

If we look at the potential $V(u)$ given in \eqref{potential}, we notice that it has three terms where the first term is positive and
the last two terms are negative definite. Therefore it has both repulsive and attractive pieces. Now we note that far away from the brane
($u \to \infty$) $G(u) \to 1$ and so $F(u) \to 1$ and thus $V(u) \to -1$. On the other hand, near the brane ($u \to 0$), $V(u) \to
-1/(4u^2) - F(u) G(u)^{-3/4}$ and since $F(u)$ is positive and the second term is negative, $V(u) \to -\infty$. So, one might think that
since in the near region the potential is attractive, gravity will not decouple from the brane. But this is not true. The point is that
in between $u=\infty$ and $u=0$, there might be some $u$ where potential can have a maximum. As we will see that this is indeed true. But
because of the complicated form of the potential it is not easy to find the value of $u$ analytically, where the maximum of potential occurs.
Below we will plot the potential $V(u)$ as a function of the radial distance from the brane $u$ in Fig. 1. In the left panel we have  
  \begin{figure}[ht]
  \begin{center}
    \includegraphics[width=0.5\textwidth,height=6cm]{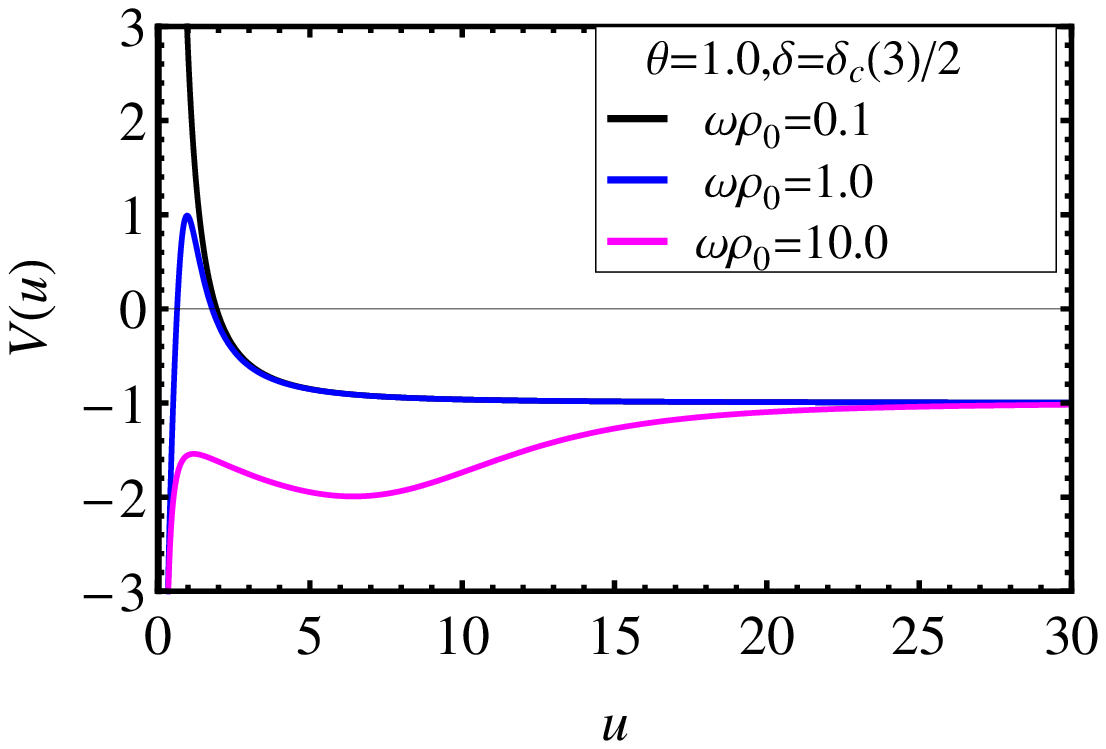}\includegraphics[width=0.5\textwidth,height=6cm]{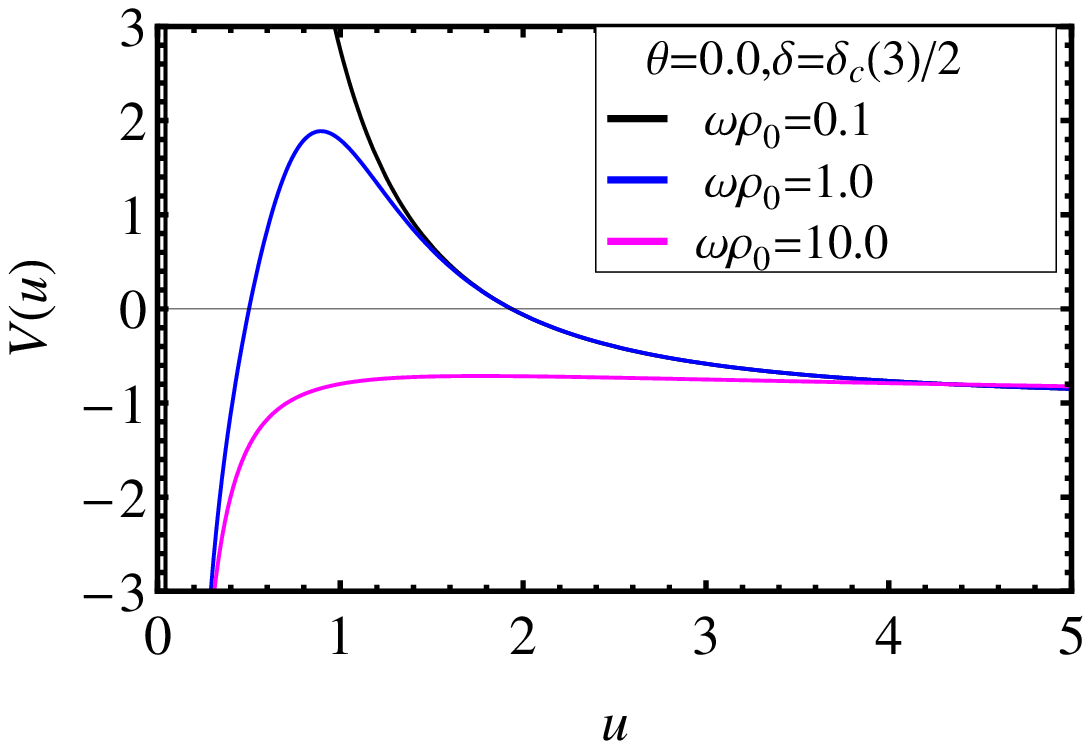}
    \caption{\it Plot of the potential $V(u)$ given in \eqref{potential} vs $u$. In the left panel the potential is given for $\theta=1.0$,
and in the right panel it is given for $\theta=0$. But in both panels $\delta=\delta_c/2=\frac{\sqrt{10}}{4}$ and $\omega\rho_0 = 10.0,\,1.0,\,0.1$.}
    \label{fig:or10}
  \end{center}
\end{figure}
plotted the potential for non-zero $\theta (=1.0)$, and so the brane is charged and in the right panel we have plotted the same for $\theta=0$,
i.e., the brane is chargeless.
Also we fixed $\delta=\delta_c/2=\sqrt{10}/4$ a typical value. In each figure we have plotted $V(u)$ for three different values of $\omega\rho_0$,
namely, 10.0, 1.0 and 0.1. Note that $\rho_0 \sim \ell_s$, where $\ell_s$ is the fundamental string length. Low energy limit or decoupling limit
means $\ell_s \to 0$. So, in this limit $\rho_0 \to 0$ and since $\omega$ is fixed $\omega\rho_0 \to 0$ in the decoupling limit. From both the
figures we find that for large $\omega\rho_0$, $V(u)$ remains negative (attractive potential) and goes to $-\infty$ as $u \to 0$. This clearly indicates that
there is no decoupling of scalar or graviton in this case. 
However as $\omega\rho_0$ decreases, there is a positive maximum of the potential in between
$u=\infty$ and $u=0$. The maximum in fact increases sharply as we further decrease $\omega\rho_0$ and eventually goes to infinity before $u$ goes
to zero. This clearly indicates that the potential acts as an infinite barrier to the scalar or the graviton in the low energy or decoupling 
limit. In other words, due to the potential of the non-supersymmetric D3 brane, the scalar or the graviton will never be able to reach the
brane in the decoupling limit and we say that gravity gets decoupled from the brane. One can also calculate the scalar or graviton absorption
cross-section by solving the equation \eqref{eom} in some situation and show that it vanishes in the decoupling limit \cite{Nayek:2015tta}.  
This is precisely what happens also for BPS D3 brane \cite{Das:1996wn,Klebanov:1997kc,Gubser:1998iu,Alishahiha:2000qf} and
this is the reason there is a theory on the brane which is non-gravitational and is dual to a gravity theory given by the geometry near the
brane. This is the origin of AdS/CFT correspondence and we see that same phenomenon also occurs for the non-supersymmetric D3 brane and gives
a strong evidence for the existence of a non-supersymmetric AdS/CFT correspondence.

We can go even a step further and find the exact decoupling limit (at least in some cases) and using that obtain the throat geometry of
non-supersymmetric D3 brane \cite{Nayek:2016hsi} in analogy with BPS D3 brane. The geometry will then represent the gravity dual of a 
(non-gravitational) QCD-like gauge theory discussed in some earlier works \cite{Constable:1999ch,Csaki:2006ji}. The decoupling limit 
we find is \cite{Nayek:2016hsi},
\bea\label{decoupling}
& & \rho=\alpha'u \nn
& & \rho_0=\alpha'u_0 \nn
& & \a \cosh^2\theta=\frac{2g_{\rm YM}^2 N}{u_0^4\alpha'^2}  = \frac{L^4}{u_0^4\alpha'^2}
\eea
along with $\a' \to 0$. Here $u$ and $u_0$ have the dimension of energy and $L^4 \sim g_{\rm YM}^2 N$ is the 't Hooft coupling, with $g_{\rm YM}^2$
is the Yang-Mills coupling and $N$ is the number of branes. The above decoupling limit reduces to BPS decoupling limit if we take the BPS limit.
To see that, in the BPS limit $u_0 \to 0$ and then $\rho_0^4 \a \cosh^2\theta \to L^4 \a'^2 = R^4$ which is precisely the BPS decoupling limit 
\cite{Aharony:1999ti}.
Note here that as $\a' \to 0$, $\theta$ goes to a very large value and since $\theta$ is related to the charge of the D3 brane by \eqref{constraints},
so, it implies that the non-supersymmetric D3 branes have large charges. So, the decoupling limit \eqref{decoupling} is valid only for large
charge D3 branes eventhough Fig.1 implies that we should have decoupling also for zero charge D3 branes. We haven't been able to figure out
the decoupling limit in this case and it remains an open problem.

Now with the above decoupling limit we find that the functions $G(\rho)$ and $F(\rho)$ change as,
\bea
&& G(\rho)\rightarrow G(u)=1+\frac{u_0^4}{u^4}\nn
&& F(\rho)\rightarrow \tilde F(u)\frac{L^4}{\a u_0^4\alpha'^2}
\eea
where, $\tilde F(u)=G^{\frac{\alpha}{2}}(u)-G^{-\frac{\alpha}{2}}(u)$. Then the non-supersymmetric D3 brane solution \eqref{nonsusyd3n} in the string frame becomes
\bea\label{throat}
&& ds^2=\alpha'\frac{L^2}{u_0^2}\left[\tilde F(u)^{-\frac{1}{2}}G(u)^{\frac{\delta}{4}}
\left(-dt^2+\sum_{i=1}^3(dx^i)^2\right)+\tilde F(u)^{\half}G(u)^{\frac{1+\delta}{4}}\left(\frac{du^2}{G(u)}+u^2d\Omega_5^2\right)\right]\nn
&& e^{2\phi}=g_s^2G(u)^\delta
\eea
Here we have restored the string coupling constant $g_s$. The Yang-Mills coupling constant is related to $g_s$ by $g_{YM}^2=2\pi g_s$ and is 
independent of $\alpha'$. Also in the above we have redefined the coordinates $(t,\, x^i) \to \frac{L^2}{\sqrt{\a} u_0^2} (t,\,x^i)$, for 
$i=1,\,2,\,3$ and rescaled $L^2 \to \sqrt{\a}L^2$.
The effective string coupling constant $e^{\phi} = \frac{g_{\rm eff}^2}{N} = g_s G(u)^{\frac{\d}{2}} = \frac{g_{\rm YM}^2}{2\pi} G(u)^{\frac{\d}{2}}$ 
is also independent of $\alpha'$. We, therefore, claim \eqref{throat} to be the throat geometry of non-supersymmetric D3 brane. It can be easily
checked that in the BPS limit $u_0 \to 0$ the above geometry reduces to AdS$_5$ $\times$ S$^5$. The same geometry can also be
obtained in the asymptotic limit, i.e., for $u \to \infty$.     

By certain coordinate transformation the above geometry \eqref{throat} can be shown to get mapped to the two parameter D3 brane solution
obtained by Constable and Myers \cite{Constable:1999ch} few years ago. They claimed this geometry to be the gravity dual of some QCD-like 
theory and our calculation
justifies this claim. In a special case the geometry above \eqref{throat} can also be shown by another coordinate transformation to be 
equivalent to the geometry studied by
Csaki and Reece \cite{Csaki:2006ji}, who claimed the geometry to be dual to some non-perturbative gauge theory and again our calculation 
justifies that claim. 

\vs{5}

{\em Acknowledgement}: I would like to thank Kuntal Nayek for collaboration on the works described here.

\end{document}